\documentclass[amsmath,amssymb,aps,pre,showpacs,showkeys]{revtex4}
\usepackage{bm,hyperref}
\newcommand{\ud}{\mathrm{d}}
\newcommand{\ue}{\mathrm{e}}
\newcommand{\als}{\alpha^*}
\newcommand{\alti}{\tilde{\alpha}}
\newcommand{\xiu}{\xi_{\textit{\scriptsize I}}}
\newcommand{\xid}{\xi_{\textit{\scriptsize II}}}
\newcommand{\xit}{\xi_{\textit{\scriptsize III}}}

\begin{document}

 \title{Fluid-particle~separation in~a~random~flow described~by~the~telegraph~model}
 \date{\today}
 \author{Gregory Falkovich}
 \author{Marco \surname{Martins Afonso}}
 \email{marcomar@fisica.unige.it}
 \affiliation{Department of Physics of Complex Systems, Weizmann Institute of Science, Rehovot (Israel)}

\begin{abstract}
 We study the statistics of the relative separation between two
 fluid particles in a spatially smooth and temporally random flow.
 The Lagrangian strain is modelled by a telegraph noise, which is
 a stationary random Markov process that can only take two values
 with known transition probabilities. The simplicity of the model
 enables us to write closed equations for the inter-particle
 distance in the presence of a finite-correlated noise. In 1D, we
 are able to find analytically the long-time growth rates of the
 distance moments and the senior Lyapunov exponent, which
 consistently turns out to be negative. We also find the exact
 expression for the Cram\'er function and show that it satisfies
 the fluctuation relation (for the probability of positive and
 negative entropy production) despite the time irreversibility of
 the strain statistics. For the 2D incompressible isotropic case,
 we obtain the Lyapunov exponent (positive) and the asymptotic
 growth rates of the moments in two opposite limits of fast and
 slow strain. The quasi-deterministic limit (of slow strain) turns
 out to be singular, while a perfect agreement is found with the
 already-known delta-correlated case.
\end{abstract}

 \pacs{47.52.+j, 47.27.-i, 47.27.eb}
 \keywords{Random flow, Batchelor regime, fluid-particle pair separation}
 \maketitle

\section{Introduction}

Understanding the statistics of the relative separation between
two fluid particles in a generic random flow is a difficult task
of significant importance \cite{O68,F63,AO91,FGV01,BF99}. Here we
treat small distances between particles, where the flow can be
considered spatially smooth (so-called Batchelor regime
\cite{B59}). We thus consider a general problem of dynamical
systems where the linearized equation for the (infinitesimal)
relative separation between two fluid particles, $\bm{R}(t)$,
takes the form
\begin{equation} \label{sigma}
 \ud_t\bm{R}(t)=\sigma(t)\bm{R}(t)\;.
\end{equation}
Here the matrix of the velocity derivatives $\sigma(t)$ is called
strain, it is taken in the co-moving (Lagrangian) frame and its
statistics is supposed to be given. At times much exceeding the
correlation time of the strain, general properties of systems with
a dynamical chaos (called Lagrangian chaos in fluid mechanics) are
described by the multi-fractal spectrum $\gamma(n)$ of the growth
rates of the moments:
\begin{equation} \label{gamma}
 \langle R^n(t)\rangle\sim\ue^{\gamma(n)t}\;.
\end{equation}
The function $\gamma(n)$ is convex, its derivative at zero is
called the senior Lyapunov exponent,
\begin{equation} \label{Lyap}
 \lambda\equiv\lim_{t\to\infty}[\langle\ln R(t)\rangle/t]=\gamma'(0)\;.
\end{equation}
Apart from that, not much is known about the general properties of
the function $\gamma(n)$. This is why few particular solvable
cases are of much importance \cite{FGV01,MK99}. For the
statistically isotropic strain, only three cases were found
amenable to an analytic treatment \cite{FGV01}: the completely
solvable short-correlated case in arbitrary dimension (where
$\gamma(n)$ is a parabola), and two partially solvable cases, the
long-correlated case in two dimensions (where the Lyapunov
exponent and its dispersion are found \cite{CFKL95}) and the
large-dimensionality case (where the second and the fourth moments
are found together with the different-time correlation functions
\cite{FKL98}).

In this paper we describe another situation which allows exact
analytical calculations: the telegraph-noise model for the strain
\cite{SL78}. This model enables one to describe analytically the
dependence of $\gamma(n)$ on the correlation time of the strain.
We shall study both the 1D (compressible) and the incompressible
2D cases. The telegraph noise is a stationary random process,
$\alpha(t)$, which satisfies the Markov property and only takes
two values, $a_1$ and $a_2$. The probability, per unit time, of
passing from the latter state to the former (or viceversa) is
given by $\nu_1$ ($\nu_2$, respectively). In what follows, we
shall consider some special cases, for which simple formulas hold.

(i) If $a_1=-a_2$, then the process itself is stochastic but its
square is deterministic, keeping the constant value
$\alpha^2(t)=a^2$ (with $a=|a_1|=|a_2|$). We shall denote such
processes with a star, e.g.~$\als(t)$.

(ii) If the average $\langle\alpha\rangle=(\nu_1a_1+\nu_2a_2)/\nu$
vanishes (with $\nu=\nu_1+\nu_2$), then the autocorrelation takes
the form
$\langle\alpha(t)\alpha(t')\rangle=\ue^{-\nu|t-t'|}(\nu_1a_1^2+\nu_2a_2^2)/\nu$.
For any analytical functional $F[\alpha]$ (function of time), a
simple ``formula of differentiation'' \cite{SL78} then holds in
this case:
\begin{equation} \label{fod}
 \ud_t\langle\alpha(t)F[\alpha]\rangle=\langle\alpha(t)\ud_tF[\alpha]\rangle-
 \nu\langle\alpha(t)F[\alpha]\rangle\;.
\end{equation}
We shall denote such processes with a tilde, e.g.~$\alti(t)$. Note
that it is always possible to reduce to this case from a generic
telegraph noise (in particular from one $\als(t)$), simply by
subtracting from the latter its mean value.

(iii) If both the properties in (i) and (ii) hold simultaneously,
the maximum level of simplification is reached. For the sake of
simplicity, we shall denote such processes not with
$\tilde{\alpha}^*(t)$ but rather with $\xi(t)$.

This type of noise has often been used to mimic more general
colored noise in stochastic theories and has also been applied,
e.g., to optics (see \cite{KWW91} and references therein). An
interesting application of the ``telegraph concept'' to turbulence
can be found in \cite{SB06}. In [10] a possible extension to the
case of a multi-valued telegraph process is discussed. While the
telegraph model is still far from any realistic flow, it is one
step closer to reality than the previous (short-correlated) models
that allowed for an analytical treatment of the two-point
separation \cite{FGV01}. Our results provide, in our opinion, a
first step in the analytic description of the role played by the
strain correlation time in the statistics of the fluid-particle
pair separation.

The paper is organized as follows: in section \ref{1d} we will
study the 1D case. In section \ref{2d} we will move to 2D,
describing the general incompressible isotropic case; the latter
will be analyzed at first in its complete form, then (respectively
in subsections \ref{erd} and \ref{qdcc}) by means of a simplified
dynamics and eventually in the short-correlated limit. The
quasi-deterministic limit is also shown in subsection \ref{app1}.
Conclusions follow in section \ref{conc}.

\section{1D case} \label{1d}

The simplest meaningful model, making use of the telegraph noise,
for the 1D case, consists in assuming a strain function with the
property (i), i.e.
\begin{equation} \label{sas}
 \sigma(t)=\als(t)\;.
\end{equation}
Indeed, the problem is compressible locally, i.e.~within every
single realization of the noise, but for fluid-mechanical
applications we want the average (1D) ``volume'' $\langle
R(t)\rangle$ to be conserved. This means that we have to impose
the constraint
\begin{equation} \label{g10}
 \gamma(1)=0\;.
\end{equation}
From (\ref{sigma}) and (\ref{sas}), it is straightforward to
obtain the system of the two first-order ordinary differential
equations which describe the time evolution of moments. It is
indeed sufficient to multiply the starting equation by the noise
itself (taken at the same time) in order to get a closed system,
thanks to the simplicity of the telegraph form. Namely, if we
define $\alpha_0\equiv\langle\als(t)\rangle$ and rewrite
accordingly $\als(t)=\alti(t)+\alpha_0$, we obtain
\begin{equation} \label{rn0}
 \ud_tR^n(t)=n\als(t)R^n(t)=n[\alti(t)+\alpha_0]R^n(t)
\end{equation}
and thus
\begin{equation} \label{rn}
 \ud_t\langle R^n(t)\rangle=n\langle\alti(t)R^n(t)\rangle+n\alpha_0
 \langle R^n(t)\rangle\;.
\end{equation}
We must then look for the evolution equation of
$\langle\alti(t)R^n(t)\rangle$, which is readily done by
exploiting (\ref{fod}):
\begin{equation} \label{arn1}
 \ud_t\langle\alti(t)R^n(t)\rangle=\langle\alti(t)\ud_tR^n(t)\rangle-\nu
 \langle\alti(t)R^n(t)\rangle\;.
\end{equation}
Taking (\ref{rn0}) into account, we get
$\langle\alti(t)\ud_tR^n(t)\rangle=na^2\langle
R^n(t)\rangle-n\alpha_0^2\langle
R^n(t)\rangle-n\alpha_0\langle\alti(t)R^n(t)\rangle$, where we
made use of the property stated in (i). Consequently, (\ref{arn1})
is rewritten as
\begin{equation} \label{arn2}
 \ud_t\langle\alti(t)R^n(t)\rangle=n(a^2-\alpha_0^2)\langle R^n(t)
 \rangle-(\nu+n\alpha_0)
 \langle\alti(t)R^n(t)\rangle\;.
\end{equation}
Equations (\ref{rn}) and (\ref{arn2}) constitute the system we
were looking for. It is easy to recast the latter as a
second-order differential equation for the quantity $\langle
R^n(t)\rangle$ alone, which can then be solved exactly:
\begin{eqnarray} \label{errenne}
 &&\ud_t^2\langle R^n(t)\rangle+\nu\ud_t\langle R^n(t)\rangle-n(na^2+ \nu\alpha_0)\langle R^n(t)\rangle=0\;,\nonumber\\
 &&\langle R^n(t)\rangle=b_n\exp\left\{-(\sqrt{\nu^2+4n^2a^2+4n\nu\alpha_0}+\nu)t/2\right\}\nonumber\\
 &&\hspace{1.8cm}+\beta_n\exp\left\{(\sqrt{\nu^2+4n^2a^2+4n\nu\alpha_0}-\nu)t/2\right\}\;,\nonumber\\
 &&\gamma(n)=(\sqrt{\nu^2+4n^2a^2+4n\nu\alpha_0}-\nu)/2\;.
\end{eqnarray}
Constraint (\ref{g10}) applied to (\ref{errenne}) implies
\begin{equation} \label{alfagamma}
 \alpha_0=-\frac{a^2}{\nu}\ \Longrightarrow\ \gamma(n)=\frac{\sqrt{\nu^2+4a^2n(n-1)}-\nu}{2}\;.
\end{equation}
This is an even function of $n-1/2$, it is convex and the
asymptotics are $\gamma(n){\sim}\pm an$ at ${n\to\pm\infty}$.

The mean value $\alpha_0$ is negative but its modulus cannot
exceed $a$, which means that only situations with $a/\nu$ (the
analogue of the Kubo number) smaller than unity are physically
relevant. Another interesting consequence is the relation
$\nu_1-\nu_2=-a$, which reduces from three to two the number of
free parameters in the original definition (\ref{sas}) of the
strain. This means that the lower level of the noise persists
longer: in the Lagrangian frame, indeed, particles spend longer
time in contracting regions than in expanding ones, thus enhancing
the weight of the former in the statistics. Accordingly, the
Lyapunov exponent, computed as the derivative of (\ref{alfagamma})
taken in the origin, turns out to be negative as is required by
the general theory \cite{R99,BFF01}:
\begin{equation} \label{lambda}
 \lambda=\gamma'(0)=-a^2/\nu\;.
\end{equation}

The joint probability density function $P(R,t)$ can also be
investigated. It satisfies the second-order partial differential
equation \cite{SL78}
\begin{equation} \label{prob}
 \partial_t^2P+\nu\partial_tP-a^2\partial_R[R\partial_R(RP)]+a^2\partial_R(RP)=0\;.
\end{equation}
Equation (\ref{prob}) has no stationary solutions, as both
$P=\textrm{const.}$ and $P\propto R^{-1}$ are non-normalizable.
The asymptotic solution has a large-deviation form
\begin{equation} \label{pro}
 P(R,t)\stackrel{t\to+\infty}{\sim}\ue^{-tH(X)}\;,
\end{equation}
where $X\equiv t^{-1}\ln[R/R(0)]$ can only belong (because of
(\ref{sigma})) to the interval $[-a,a]$. The Cram\'er function
$H(X)$ is simply the Legendre transform of $\gamma(n)$:
\begin{equation} \label{cramer}
 H(X)=\frac{\nu}{2}+\frac{X}{2}-\frac{\sqrt{(\nu^2-a^2)(a^2-X^2)}}{2a}\;.
\end{equation}
It is a convex function, which exactly satisfies the
Evans--Searles--Gallavotti--Cohen fluctuation relation
$H(X)-H(-X)=X$ \cite{ES94,GC95} despite the fact that the strain
statistics is not time-reversible \cite{FF04}. The function
vanishes quadratically at the minimum, represented by the Lyapunov
exponent,
$H(X)\stackrel{X\sim\lambda}{\sim}(X-\lambda)^2\nu^3/4a^2(\nu^2-a^2)$,
and it approaches vertically the boundaries of its compact domain:
\[H(X)\stackrel{X\sim\pm a}{\sim}\frac{\nu\pm a}{2}-\sqrt{\frac{\nu^2-a^2}{2a}}\sqrt{a\mp X}\;.\]
Plugging (\ref{cramer}) into (\ref{pro}), we obtain the final
result:
\[P(R,t)\stackrel{t\to+\infty}{\sim}|R|^{-1/2}{\exp\left\{\left(\sqrt{(1-a^2/\nu^2)\{1-\ln^2[R/R(0)]/a^2t^2\}}-1\right)\nu t/2\right\}}\;.\]

It is worth mentioning what happens if one does not impose the
constraint (\ref{g10}) and assumed a strain $\sigma(t)=\xi(t)$
with the property described in (iii). In this case,
$\gamma(n)=(\sqrt{\nu^2+4n^2a^2}-\nu)/2$ and the Lyapunov exponent
would vanish. The Cram\'er function would be
$H(X)=\nu/2-\nu\sqrt{a^2-X^2}/2a$, so that
\[P(R,t)\stackrel{t\to\infty}{\sim}|R|^{-1}\exp\left[-\left(1-\sqrt{1-\ln^2[R/R(0)]/a^2t^2}\right) \nu t/2\right]\]
would be the solution to the equation for the joint probability
distribution, similar to (\ref{prob}) but without the last term.

\section{2D case} \label{2d}

In the 2D case, it is possible to consistently impose the
incompressibility constraint locally, for every single realization
of the noise. Since this case is most interesting for
applications, we shall only consider this situation, which implies
the strain matrix to be traceless.

Consider a general statistically isotropic flow described by the
strain matrix
\[\sigma_{ij}(t)=\left(\begin{array}{cc}
 \xiu(t)&\xid(t)+\sqrt{2}\xit(t)\\
 \xid(t)-\sqrt{2}\xit(t)&-\xiu(t)
\end{array}\right)\]
The noises $\xiu(t)$, $\xid(t)$ and $\xit(t)$ are independent of
one another but share the properties in (iii) with the same
coefficients $a$ and $\nu$, which can now range from $0$ to
$\infty$ independently. Differently from the 1D case, we will see
that this is enough to mimic a realistic situation. The matrix
$\sigma$ thus turns out to be singular and nilpotent:
$\det\sigma=0=\sigma^2$.\\
By means of simple manipulations, analogue to those described for
the 1D case, one can write down a closed system for the moments of
the components $r_{n,k}(t)\equiv R_1^k(t)R_2^{n-k}(t)$ at every
$n$, in the form
\begin{equation} \label{ra}
 \ud_t\langle\mathcal{R}^{(n)}_{\iota}\rangle=\mathtt{A}^{(n)}_{\iota\kappa}
 \langle\mathcal{R}^{(n)}_{\kappa}\rangle\;.
\end{equation}
The column vector $\mathcal{R}^{(n)}$ has dimension $8(n+1)$,
being made up of: the $n+1$ components of $r_{n,\bullet}$ itself,
the $3(n+1)$ components of $r_{n,\bullet}$ times one noise, the
$3(n+1)$ components of $r_{n,\bullet}$ times two (distinct)
noises, and finally the $n+1$ components of $r_{n,\bullet}$ times
all the three noises. No other quantity is required to close the
system, because as soon as a noise appears as square it becomes a
deterministic quantity and can be taken out of the statistical
average. The matrix $\mathtt{A}^{(n)}$ is thus of order
$8(n+1)\times8(n+1)$ and turns out to be dependent only on $a$ and
$\nu$. The number of noises sharply increases in higher
dimensions, thus strongly enhancing the number of components in
$\mathcal{R}^{(n)}$: this is why we confine ourselves to the 2D
case.

To study the long-time evolution of $\langle r_{n,k}\rangle$, one
should identify the eigenvalue of $\mathtt{A}^{(n)}$ with the
largest positive real part. This is not an easy task from the
numerical point of view for growing $n$, and it does not look
feasible analytically for a generic $n$, even if the matrix itself
can be easily written down as a function of $n$ (the details can
be found in \cite{MMA} where, in particular, the whole spectrum of
eigenvalues is found for $n=1$ and $n=2$).

The evolution of the linear components $\langle R_1\rangle$ and
$\langle R_2\rangle$ is straightforward to describe: the largest
eigenvalue is $0$, which means that they do not show any
exponential growth at large times; however, as they can change
sign, this does not yield any information on $\langle R\rangle$,
i.e.~on $\gamma(1)$. The situation is more interesting for the
quadratic components ($R_1^2$, $R_1R_2$, $R_2^2$), because in this
case the largest eigenvalue is given by
\begin{equation} \label{mu}
\gamma(2)=-\nu+\frac{{\nu}^2}{\sqrt[3]{216a^2\nu+3\sqrt{5184a^4\nu^2-3\nu^6}}}+
 \frac{\sqrt[3]{216a^2\nu+3\sqrt{5184a^4\nu^2-3\nu^6}}}{3}\;,
\end{equation}
which is always positive (i.e. describes an exponential growth of
$\langle R^2\rangle=\langle R_1^2\rangle+\langle R_2^2\rangle$)
and behaves asymptotically as
\begin{equation} \label{as}
\gamma(2)=\left\{\begin{array}{ll}
  \displaystyle8\frac{a^2}{\nu}-96\frac{a^4}{\nu^3}\left[1+O\left(\frac{a}
  {\nu}\right)^2\right]&\displaystyle\textrm{for }\frac{a}{\nu}\to0\\[0.3cm]
  \displaystyle2\sqrt[3]{2a^2\nu}-\nu\left[1+O\left(\frac{\nu}{a}\right)^{2/3}
  \right]&\displaystyle\textrm{for }\frac{a}{\nu}\to\infty\;.
 \end{array}\right.
\end{equation}

\subsection{Exact reduced dynamics} \label{erd}

Since the general process is computationally very demanding, we
succeeded to find a reduced set of variables invariant with
respect to the evolution, i.e.~a subset of components of
$\mathcal{R}^{(n)}$ such that the corresponding rows in
$\mathtt{A}^{(n)}$ are nonzero only in columns whose index has
been taken into account in the reduced subset. Consider even $n$:
\begin{equation} \label{binom}
 R^n=(R_1^2+R_2^2)^{n/2}=\sum_{k=0,2,\ldots,n}\binom{n/2}{k/2}r_{n,k}\,,
\end{equation}
by definition for positive even $n$, thus the values $\gamma(n)$
can be extracted from the knowledge of $n/2+1$ of the first $n+1$
components in $\mathcal{R}^{(n)}$. On the contrary, this is not
the case for the odd $n$'s, which behave similarly to non-natural
$n$'s because they involve nonlinear operations (in this case a
square root) that do not commute with the statistical
averaging.

In particular, we analyzed the values $n=4$, $n=6$ and $n=8$.
Adopting a technique described in \cite{MMA}, one gets matrices of
order 5, 7 and 9 respectively (i.e.~$n+1$), whose eigenvalues with
largest positive real part behave asymptotically as:
\[\gamma(4)\sim\left\{\begin{array}{ll}
 \displaystyle24\frac{a^2}{\nu}-576\frac{a^4}{\nu^3}&\displaystyle\textrm{for }\frac{a}{\nu}\to0\\[0.3cm]
 \displaystyle2\sqrt[5]{36a^4\nu}&\displaystyle\textrm{for }\frac{a}{\nu}\to\infty\;,
\end{array}\right.\]
\[\gamma(6)\sim\left\{\begin{array}{ll}
 \displaystyle48\frac{a^2}{\nu}-2016\frac{a^4}{\nu^3}&\displaystyle\textrm{for }\frac{a}{\nu}\to0\\[0.3cm]
 \displaystyle2\sqrt[7]{1800a^6\nu}&\displaystyle\textrm{for }\frac{a}{\nu}\to\infty\;,
\end{array}\right.\]
\[\gamma(8)\sim\left\{\begin{array}{ll}
 \displaystyle80\frac{a^2}{\nu}-5280\frac{a^4}{\nu^3}&\displaystyle\textrm{for }\frac{a}{\nu}\to0\\[0.3cm]
 \displaystyle2\sqrt[9]{181440a^8\nu}&\displaystyle\textrm{for }\frac{a}{\nu}\to\infty\;.
\end{array}\right.\]
These relations provide us with three points of the curve
$\gamma(n)$ in the asymptotic conditions, to which three more can
be added: besides the already-mentioned $\gamma(2)$, one should
indeed remember that both $\gamma(0)$ and $\gamma(-2)$ (in 2D)
must vanish \cite{F63,ZRMS84,FGV01}. A simple extrapolation then
suggests the following asymptotic behaviour:
\begin{equation} \label{ggg}
 \gamma(n)\sim\left\{\begin{array}{ll}
  \displaystyle G(n)\frac{a^2}{\nu}+\mathcal{G}(n)\frac{a^4}{\nu^3}
  &\displaystyle\textrm{for }
  \frac{a}{\nu}\to0\\[0.3cm]
  \displaystyle g(n)a^{n/(n+1)}\nu^{1/(n+1)}&\displaystyle\textrm{for }
  \frac{a}{\nu}\to\infty\;.
 \end{array}\right.
\end{equation}
Note that the noise correlation time is $T=1/\nu$. The $n$-th
growth rate grows linearly with $T$ when the latter is small and
decays as $T^{-1/(n+1)}$ when it is large.

The coefficient $G(n)$ can be proven rigorously to be $n(n+2)$
(see the next subsection), while such a proof does not look
feasible for $\mathcal{G}(n)$. However, starting from the
quadratic form of the former, one can guess a fourth-order
polynomial for the latter, $\mathcal{G}(n)=3n(n+2)(n^2+2n+8)/4$,
and the correctness of this expression seems to be confirmed by
the fact that it satisfies the six conditions in $n=-2,0,2,4,6,8$
despite possessing only five degrees of freedom. As the following
coefficients in the power expansion at small $a/\nu$ are expected
to show higher powers of $n$, such an expansion is not uniform in
the sense that it must fail for some (large enough) value of $n$.
On the contrary, the coefficient $g(n)$ must originate from some
combinatorics, but has not been identified yet, while the presence
of non-integer powers of $a$ and $\nu$ suggests that the limit
${a}/{\nu}\to\infty$ is singular. From (\ref{ggg}), exploiting the
knowledge of $G(n)$ and the vanishing of $g(0)$, we get the
asymptotic behaviour of the Lyapunov exponent:
\begin{equation} \label{lam}
 \lambda=\left\{\begin{array}{ll}
  \displaystyle2\frac{a^2}{\nu}&\displaystyle\textrm{for }\frac{a}{\nu}\to0\\[0.3cm]
  \displaystyle g'(0)\nu&\displaystyle\textrm{for }\frac{a}{\nu}\to\infty\;.
 \end{array}\right.
\end{equation}

\subsection{Short-correlated case} \label{qdcc}

The results obtained in the previous subsection are exact, but
they have been obtained for particular values of $n$, and the
generalization of the coefficients (\ref{ggg}) as functions of $n$
was empirical. There is however at least one limit in which
rigorous analytical results can be obtained: it is provided by the
short-correlated case, corresponding to small $a/\nu$. We remind
that in the exactly delta-correlated case \cite{FGV01,BF99} the
probability distribution of $R(t)$ is lognormal and the Cram\'er
function is quadratic. Note that the correct limit is such that
$a/\nu\to0$ but $a^2/\nu$ is finite. The key point in this limit
is that the leading behaviour can be extracted rigorously, for any
natural $n$, from a restricted dynamics, namely taking into
account only the first $4(n+1)$ components in $\mathcal{R}^{(n)}$
(corresponding to the average of the coordinates $r_{n,k}$, and of
the latter times one noise). The details of the calculations can
be found in \cite{MMA}; here we only mention that, upon
introducing the small-Kubo Taylor expansion, we face a
\emph{degenerate perturbation theory}. Indeed, the eigenvalues
corresponding to a fixed $n$ are not definitely separate for
$a/\nu\to0$, but rather some of them vanish in this limit, and we
have to identify the one which keeps the largest positive real
part in this process. The result of this analysis is that in the
evolution equation for $\langle r_{n,k}\rangle$ one should account
for the complete form, while for
$\langle\xi_{\bullet}r_{n,k}\rangle$ one can neglect all source
terms involving two different noises: the system thus closes at
this stage, reducing the order of the corresponding matrix from
the initial $8(n+1)$ to $4(n+1)$. In other words, only the
upper-left quarter of the matrix $\mathtt{A}^{(n)}$ is relevant.
The problem can now easily be recast as a system of $n+1$
second-order differential equations for $r_{n,k}$. Indeed, we have
\begin{eqnarray*}
 \ud^2_tr_{n,k}&=&kR_1^{k-1}R_2^{n-k}\ud^2_tR_1+(n-k)R_1^kR_2^{n-k-1}\ud^2_tR_2\\
 &&+k(k-1)R_1^{k-2}R_2^{n-k}(\ud_tR_1)^2+(n-k)(n-k-1)R_1^kR_2^{n-k-2}(\ud_tR_2)^2\\
 &&+2k(n-k)R_1^{k-1}R_2^{n-k-1}\ud_tR_1\ud_tR_2\;,
\end{eqnarray*}
and, keeping into account the aforementioned simplification in the
equations for $\langle\xi_{\bullet}r_{n,k}\rangle$,
\begin{eqnarray} \label{secord}
 \ud^2_t\langle r_{n,k}\rangle+\nu\ud_t\langle r_{n,k}\rangle&=&a^2\left[(n^2+6k^2-6nk-n)\langle r_{n,k}\rangle\right.\\
 &&+3k(k-1)\langle r_{n,k-2}\rangle+3(n-k)(n-k-1)\langle r_{n,k+2}\rangle\big]\nonumber\\
 &\equiv&a^2\mathsf{M}^{(n)}_{kk'}\langle r_{n,k'}\rangle\;.\nonumber
\end{eqnarray}
The matrix $\mathsf{M}^{(n)}$, defined by the right-hand side of
(\ref{secord}), is a priori of order $(n+1)\times(n+1)$. However,
it is apparent from (\ref{secord}) that, for a fixed $n$, the
dynamics of the components of $\langle r_{n,k}\rangle$ with even
$k$ are independent of those with odd $k$. Therefore, as our main
goal is to reconstruct objects like (\ref{binom}), we can just
focus on subset $\hat{\mathsf{M}}^{(n)}$ of $\mathsf{M}^{(n)}$
consisting of the lines with even $k$, and thus reduce to a
$(n/2+1)\times(n/2+1)$ problem, simply neglecting the lines
corresponding to odd $k$. In other words, we rewrite
(\ref{secord}) as
\begin{equation} \label{hat}
 \ud^2_t\langle r_{n,2k}\rangle+\nu\ud_t\langle
 r_{n,2k}\rangle=a^2\sum_{k'=0,2,\ldots,n}\mathsf{M}^{(n)}_{2k,k'}\langle
 r_{n,k'}\rangle=a^2\hat{\mathsf{M}}^{(n)}_{kk'}\langle r_{n,2k'}\rangle\;.
\end{equation}
Moreover, it is easy to show that the largest-positive-real-part
eigenvalue $\mu_n$ (divided by $a^2/\nu$) that we are looking for
must belong to the spectrum of the subset matrix
$\hat{\mathsf{M}}^{(n)}$. Such subset matrix is tri-diagonal
central-symmetric but not symmetric, which means that left and
right eigenvectors differ. It is a simple task to prove that the
value $n(n+2)$ (guessed for $\mu_2/(a^2/\nu)$ asymptotically from
the previous subsection) is actually an eigenvalue, but it is not
possible to use a variational method to show that it has the
largest positive real part. However, this is not needed, because
the row vector built with the $n/2+1$ components of the binomial
coefficient $\binom{n/2}{k}$ (for $k=0,1,\ldots,n/2$) happens to
be the left eigenvector of $\hat{\mathsf{M}}^{(n)}$ corresponding
to the eigenvalue $n(n+2)$. Therefore, using (\ref{binom}) and
(\ref{hat}), one has (for details see \cite{MMA}):
\[\ud^2_t\langle R^n\rangle+\nu\ud_t\langle R^n\rangle=n(n+2)a^2\langle R^n\rangle\;.\]
This implies
\[\mu_n\stackrel{a/\nu\to0}{\sim}\frac{\sqrt{\nu^2+4n(n+2)a^2}-\nu}{2}=n(n+2)\frac{a^2}{\nu}\left[1+O\left(\frac{a}{\nu}\right)^2\right]\]
as expected. Notice that the next-leading correction cannot be
captured at this stage, because of the approximation introduced
previously. To take it into account correctly, one should
reformulate the analysis performed in this section, including also
quantities like $\langle\xi_{\bullet}\xi_{\circ}r_{n,k}\rangle$
but however excluding $\langle\xiu\xid\xit r_{n,k}\rangle$,
i.e.~considering a reduced dynamics in the first $7(n+1)$
components of $\mathcal{R}^{(n)}$.

It is worth mentioning that an analysis like the previous one, but
applied to odd $n$'s, leads to
\[\mu_n\stackrel{a/\nu\to0}{\sim}(n-1)(n+3)\frac{a^2}{\nu}\;,\]
in accordance e.g.~with the exact result $\mu_1=0$ found with the
complete dynamics. However, as already pointed out, such values
are not related to the curve $\gamma(n)$.

\subsection{Deterministic limit in 2D} \label{app1}

The deterministic case corresponds to $\nu=0$, when the noise
$\xi(t)$ takes a constant value $\pm a$. In 2D, one gets
$\ud^2_tR_1=0=\ud^2_tR_2$ (for any of the $2^3=8$ possible
combinations of the noise signs), thus both components are linear
in time and $R^n\stackrel{t\to\infty}{\sim}t^n$. Consequently, a
power-law temporal dependence is found for the separation and
$\lambda=0$. It is worth noticing that this result is due to the
fact that, with our choice, $\sigma^2=0$, and is characteristic of
the 2D isotropic situation. Therefore the limit is singular: an
exponential growth is expected in general for finite $a/\nu$, but
a power law is found when this ratio is infinite. As a result,
perturbation theory and expansion in integer powers of $\nu/a$
does not work. In (\ref{ggg}) this is reflected by the presence of
non-integer powers and the difference between the limits
$a\to\infty$ and $\nu\to0$.

\section{Conclusions} \label{conc}

By assuming the telegraph-noise model for the velocity gradient
(or strain matrix), we were able to carry out analytical
computations and to obtain several results on the separation
between two fluid particles. Focusing on smooth flows (Batchelor
regime), we firstly analysed the one-dimensional compressible
case, finding explicit expressions for the long-time evolution of
the interparticle-distance moments (\ref{alfagamma}), the Lyapunov
exponent (\ref{lambda}) and the Cram\'er function (\ref{cramer}).
Then we concentrated on the two-dimensional incompressible
isotropic case, where a thorough analysis on the complete dynamics
was made for the evolution of linear ($n=1$) and quadratic ($n=2$)
components. Due to high computational cost, at higher $n$ we
focused on a restricted, though exact, dynamics: however, only
specific values of $n$ could be studied in this way, leading to
the extrapolations (\ref{ggg}) and (\ref{lam}). Such guess was
rigorously proved in the quasi-delta-correlated limit, for which
approximated equations were introduced.

We believe that the present paper represents an interesting
example of the use of a coloured noise for which the closure
problem can be solved analytically. In particular, we established
the dependence of the distance growth rates and of the Lyapunov
exponents on the correlation time.

\begin{acknowledgments}
 We thank Itzhak Fouxon, Vladimir Lebedev, Stefano Musacchio,
 Konstantin Turitsyn and Marija Vucelja for useful discussions and
 suggestions.
\end{acknowledgments}

\bibliography{AfonsoX2}

\end{document}